\documentclass[journal]{IEEEtran}
\usepackage[utf8]{inputenc}
\usepackage[T1]{fontenc}
\usepackage{algorithm}
\usepackage{amsmath}
\usepackage{amsthm}
\usepackage{amssymb}
\usepackage{graphicx}
\usepackage{setspace}
\usepackage{epstopdf}
\usepackage[font={normal}]{caption}
\usepackage{cite}
\usepackage{bm}
\usepackage{algpseudocode}
\usepackage{balance}
\usepackage{array,multirow,makecell}\setcellgapes{1pt}
\usepackage{comment}
\usepackage[table]{xcolor}
\usepackage{setspace}
\usepackage{subfigure}

\newcolumntype{R}[1]{>{\raggedleft\arraybackslash }b{#1}}
\newcolumntype{L}[1]{>{\raggedright\arraybackslash }b{#1}}
\newcolumntype{C}[1]{>{\centering\arraybackslash }b{#1}}
\makegapedcells

\begin{document}

\title{Federated Learning over Energy Harvesting Wireless Networks}
\author{Rami~Hamdi,~\IEEEmembership{Member,~IEEE,}
	Mingzhe~Chen,~\IEEEmembership{Member,~IEEE,}
	Ahmed~Ben~Said,	
	Marwa~Qaraqe,~\IEEEmembership{Member,~IEEE,}
	and~H.~Vincent~Poor,~\IEEEmembership{Fellow,~IEEE}
\thanks{A preliminary version of this work~\cite{gc} was submitted in the 2021 IEEE Global Communications Conference.} 
\thanks{R. Hamdi and M. Qaraqe are with the Division of Information and Computing Technology, College of Science and Engineering, Hamad Bin Khalifa University, Qatar Foundation, Doha, Qatar (email: hrami@hbku.edu.qa; mqaraqe@hbku.edu.qa).}
\thanks{ M. Chen and H. V. Poor are with the Department of Electrical Engineering, Princeton University, Princeton, NJ, 08544, USA (email: mingzhec@princeton.edu; poor@princeton.edu).}
\thanks{A. B. Said is with the Computer Science and Engineering Department, College of Engineering, Qatar University, Doha, Qatar (email: abensaid@qu.edu.qa).}
\thanks{Copyright (c) 2021 IEEE. Personal use of this material is permitted. However, permission to use this material for any other purposes must be obtained from the IEEE by sending a request to pubs-permissions@ieee.org.}}
\maketitle
\thispagestyle{empty}

\begin{abstract}
In this paper, the deployment of federated learning (FL) is investigated in an energy harvesting wireless network in which the base station (BS) employs massive multiple-input multiple-output (MIMO) to serve a set of users powered by independent energy harvesting sources. Since a certain number of users may not be able to participate in FL due to the interference and energy constraints, a joint energy management and user scheduling problem in FL over wireless systems is formulated. This problem is formulated as an optimization problem whose goal is to minimize the FL training loss via optimizing user scheduling. To find how the factors such as transmit power and number of scheduled users affect the training loss, the convergence rate of the FL algorithm is first analyzed. Given this analytical result, the user scheduling and energy management optimization problem can be decomposed, simplified, and solved. Further, the system model is extended by considering multiple BSs. Hence, a joint user association and scheduling problem in FL over wireless systems is studied. The optimal user association problem is solved using the branch-and-bound technique. Simulation results show that the proposed user scheduling and user association algorithm can reduce training loss compared to a standard FL algorithm.
\end{abstract}


\section{Introduction}
Machine learning (ML) is a powerful tool that will play a critical role in designing and optimizing 6G architectures, protocols, and operations. In particular, ML enables future networks to support new costly services such as autonomous driving, enhanced mobile broadband, ultra reliable and low-latency communications, and intelligently manage wireless resource and devices~\cite{new1,new2}. However, centralized ML techniques are becoming increasingly costly, due to the increased amount of data collected through wireless edge users, the limited power and bandwidth available, and privacy concerns. Federated learning (FL) allows devices to train ML models without data transmission. This is done by having each edge user train and build its own ML model and send its trained ML model parameters to a central data center for aggregation so as to generate a shared ML model~\cite{fl0,fl1}. The aggregated model parameters are then sent to the edge users for updating their individual ML models. This procedure is repeated in several communication rounds until achieving an acceptable accuracy of the trained model. Since all data is kept on the edge users, FL can also improve data privacy and security~\cite{fl3}. The applicability of FL to edge computing and caching networks has been demonstrated in~\cite{fl4} where FL allows prediction of content popularity based on user-content interaction. However, the implementation of FL is challenging in wireless networks, due to multiuser interference, energy constraints, and heterogeneity of the users in terms of data model and hardware, such as CPU and memory. Moreover, since a massive number of edge users will potentially participate in FL and the wireless bandwidth is limited, communication is a critical bottleneck in FL systems~\cite{fl5}. Thus, to deploy FL over wireless networks, the problem of resource allocation (i.e. user scheduling, resource block allocation, power allocation, and bandwidth splitting) must be addressed.
\vspace{-0.45cm}
\subsection{Related Work} 
Recently, several techniques have been proposed to overcome issues in FL systems such as local updating and model compression. The impact of wireless channel randomness on the FL performance was studied in~\cite{chan1} where the authors proposed a novel efficient transmission method based on gradient sparsification. Moreover, the impact of noise on the FL loss function was investigated in~\cite{chan2} and a robust transmission scheme was proposed to eliminate the effect of noise. The incorporation of FL into a more sophisticated wireless system based on cell-free massive multiple-input multiple-output (MIMO) was considered in~\cite{chan3} and a transmission scheme was developed to optimize FL performance. A number of existing works~\cite{us1,us2,us3,us4,us5,us6,us7,us8,us9,us10,us11} have studied user scheduling in FL over wireless networks under various assumptions and network architectures. The authors in~\cite{us1} investigated various user scheduling policies to enhance FL performance. In~\cite{us2}, the authors investigated the problem of user selection in FL over wireless systems and proposed an efficient user scheduling scheme under wireless resource constraints. The authors of~\cite{us3} minimized the convergence time of FL by designing an appropriate user selection scheme. In~\cite{us4}, the access probabilities of users were optimized for given local updates considering a multichannel random access. In~\cite{us5}, the authors optimized user selection in FL training in order to avoid collisions and to minimize transmission failures. An efficient data sampling and user selection algorithm was designed in~\cite{us6} to enhance FL performance in mobile computing systems. The authors of~\cite{us7} overcame the issue of imbalanced data in FL over mobile systems by proposing an adaptive user scheduling and data augmentation algorithm. In~\cite{us8}, a joint user scheduling and resource block allocation scheme was proposed to minimize the FL training loss over wireless systems assuming imperfect channel state information (CSI). In~\cite{us9}, an optimal user selection scheme was proposed to minimize the FL convergence rate. In~\cite{us10}, the authors discussed the reliability of FL performance where only a subset of users are selected to eliminate the impact of unreliable users, thereby improving the FL training accuracy. In~\cite{us11}, the authors proposed a user selection scheme to mitigate the straggler effect for FL in cell-free massive MIMO systems.

The energy efficiency of FL over wireless networks has been studied in~\cite{ener1,ener2,ener3}. In particular, the authors in~\cite{ener1} studied the energy efficiency of FL while considering limited energy budgets of wireless users and proposed an efficient iterative resource allocation algorithm. Meanwhile, power allocation among users was investigated in~\cite{ener2} by considering federated cooperation to enhance FL performance. An energy-efficient bandwidth allocation and scheduling algorithm was developed in~\cite{ener3} to reduce energy consumption of users in FL over wireless networks. In addition, a wireless FL framework was proposed in~\cite{res1} by formulating an optimization problem that captures the trade-off between communication and computation latencies, and a closed-form optimal solution was derived. In~\cite{res2}, the authors analyzed the trade-off between local updates and global updates in terms of convergence time by proposing an efficient control algorithm. Further, the authors of~\cite{res3} proposed an efficient algorithm that optimizes the convergence rate of FL under a total training time budget. The fairness in terms of comparable performance across the users in FL over wireless networks was investigated in~\cite{res4} by proposing data importance aware resource management schemes. FL is also studied for optimizing wireless network performance. In particular in~\cite{uav2}, FL is used to estimate distribution of the queue lengths in unmanned aerial vehicle (UAV) based wireless networks. In~\cite{uav3}, a selective model aggregation algorithm was proposed to enhance the accuracy of FL over UAVs wireless networks. However, implementing FL over distributed users increases the power consumption of users. Most of these existing works~\cite{us1,us2,us3,us4,us5,us6,us7,us8,us9,us10,us11,ener1,ener2,ener3,res1,res2,res3,res4} did not consider the use of energy harvesting for the implementation of FL. Powering the devices by energy harvesting sources allows them to continually acquire energy from nature or man-made phenomena (e.g., solar, wind, and electromagnetic energy). In particular, each device can use the harvested energy for FL parameter update and transmission thus increasing the number of devices that can participate in FL and improving FL convergence time and training loss.\\

This work proposes to incorporate energy harvesting into FL over wireless systems enabled by massive MIMO in order to enhance the energy efficiency by assuming that each user is powered by its own energy harvesting source. The design of energy-efficient FL over wireless systems powered by energy harvesting is challenging due to the intermittency  of renewable energy sources.\\
\\
\vspace{-1cm}
\subsection{Contributions} 
The main contribution of this work is to develop a novel energy harvesting FL framework. To our best knowledge, \emph{this is the first work that considers the implementation of FL over energy harvesting wireless networks}. The key contributions are summarized as follows:
\begin{itemize}
\item A novel energy harvesting FL framework is investigated where the base station (BS) is equipped with a massive MIMO system and each user is powered by its own energy harvesting source. Hence, a joint energy management and user scheduling problem is formulated.
\item To analyze how wireless factors such as transmit power and the number of scheduled users affect FL performance, the convergence rate of the FL algorithm is analytically investigated. 
\item Given the relationship between user scheduling and the convergence rate of the FL algorithm, the original optimization problem is decomposed and simplified, which allows us to analytically solve it. 
\item Further, the proposed system is extended to a system that consists of multiple BSs. For a multiple BS based system, one must determine the user association with different BSs and user scheduling. The relationship between the convergence rate of the FL algorithm and user association is analytically investigated.
\item Given the relationship between user association and the convergence rate of the FL algorithm, the user association problem is simplified and can be solved using branch and bound approaches.
\end{itemize}
Simulation results show the efficiency of the proposed resource management schemes in FL over energy harvesting wireless systems in terms of accuracy.

The rest of this paper is organized as follows. The FL system model and problem formulation are described in Section II. The impact of wireless parameters on the FL performance is investigated in Section III. The optimal resource allocation solution is derived in Section IV. The multiple BS system is investigated in Section V. Numerical results are presented and discussed in Section VI. Conclusions are drawn in Section VII. 

\begin{table*}[t]
	\centering
	\caption{Summary of Important Notations.}
	\label{tab1}
	\begin{tabular}{|C{2.9cm}|L{4.4cm}||C{1.5cm}|L{5.7cm}|}
		\hline   \textbf{Symbol} & \textbf{Description} & \textbf{Symbol} & \textbf{Description}\\
		\hline   $N$ & Number of antennas at the BS & $\lambda$ & Learning rate \\
		\hline   $K$ & Number of users & $\boldsymbol{h}_k(i)$ & Small-scale fading channel vector of user $k$ at frame $i$ \\
		\hline   $L$ & Number of frames & $\beta_k$ &  Path loss of user $k$ \\
		\hline   $\mathcal{D}_k$ & Dataset of user $k$ & $\boldsymbol{g}_k(i)$ & Channel vector of user $k$ at frame $i$ \\
		\hline   $M_k$ &  Number of the samples collected by user $k$ & $p_k(i)$ & Power allocated to user $k$ at frame $i$ \\
		\hline   $M$ & Total number of training data samples & $\gamma_k(i)$ & SINR for user $k$ at frame $i$ \\
		\hline   $\boldsymbol{w}_k(i)$ & Weight vector of the local FL model of user $k$ at frame $i$ & $B_{\textrm{max}}$ & Maximal battery capacity \\
		\hline   $\gamma_{th}$ &  Minimum required SINR & $E_{c}$ &  Energy consumption of user $k$ \\
		\hline   $\chi_k(i)$ &  User scheduling index of user $k$ at frame $i$ & $B_k(i)$ & Battery level of user $k$ at frame $i$ \\
		\hline   $\boldsymbol{\chi}(i)$ &  Vector of user scheduling indexes at frame $i$ & $E_k(i)$ & Amount of harvested energy for user $k$ at frame $i$ \\
		\hline   $\boldsymbol{q}$ &  Global model & $\mathcal{N}_{1}(i)$ & Set of scheduled users at frame $i$ \\
		\hline   $\boldsymbol{x}_{k,m}$ & FL input of user $k$ & $\mathcal{N}_{2}(i)$ & Set of unscheduled users at frame $i$ \\
		\hline   ${y}_{k}$ &  FL output of user $k$ & $S$ & Number of BSs \\	
		\hline   $f(\boldsymbol{w}_k(i),\boldsymbol{x}_{k,m},{y}_{k})$ & Loss function & $\alpha_{k,s}$ & User association index between user $k$ and BS $s$ \\
		\hline
	\end{tabular}
\end{table*}

\section{System Model and Problem Formulation}
\subsection{Federated Learning Model}
\begin{figure}[t]
	\centerline{\includegraphics [width=1.05\columnwidth]  {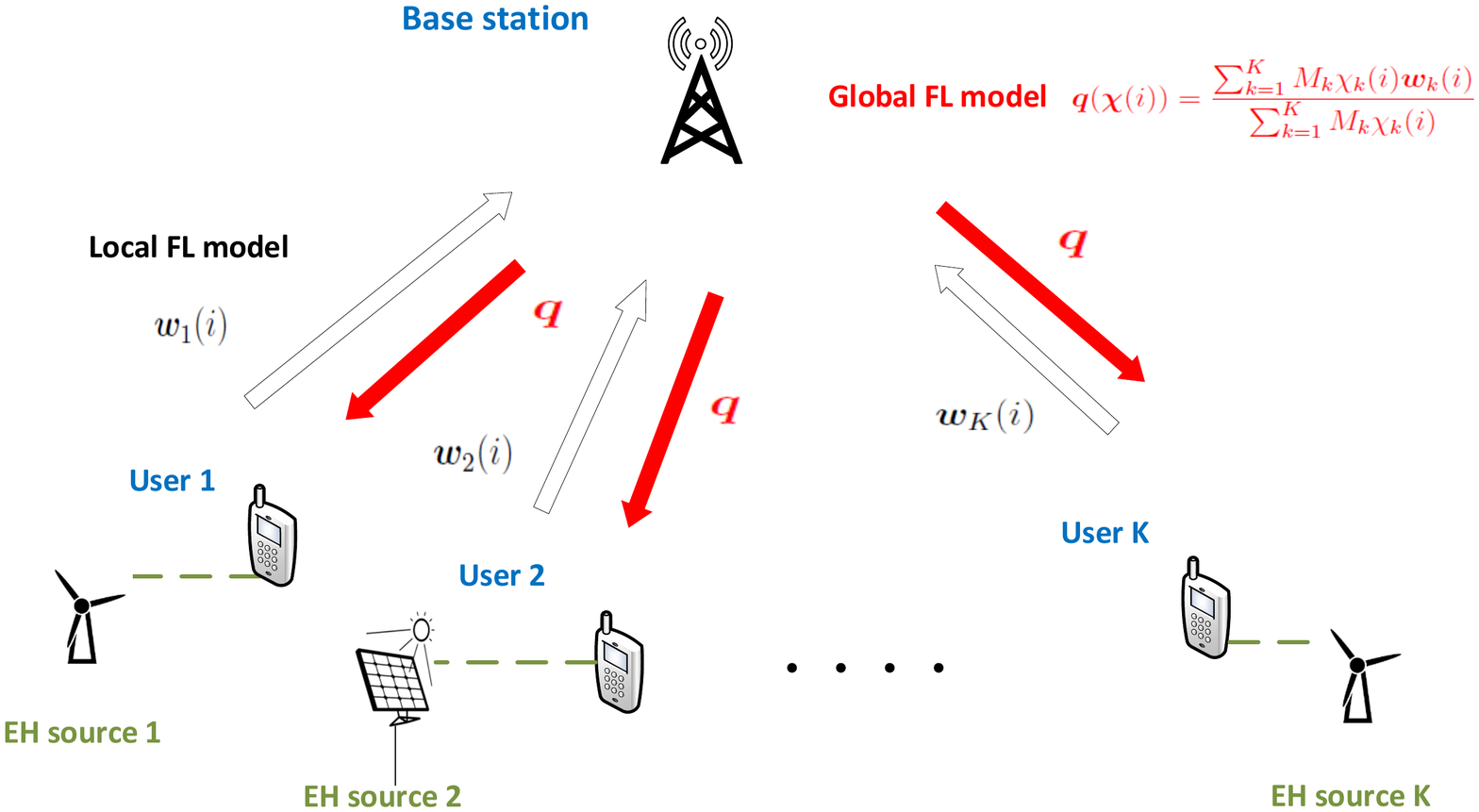}}
	\caption{Architecture of federated learning over an energy harvesting wireless network.}
	\label{fig1}
\end{figure}

Massive MIMO is a key technology for next generation wireless networks since high data rate can be achieved when large antenna arrays are adopted~\cite{mimo}. Hence, we consider an FL based wireless communication system that consists of a BS equipped with a massive MIMO system of $N$ antennas serving $K$ arbitrarily distributed single-antenna users with $N \gg K$ as shown in Fig.~\ref{fig1}. A given time interval is partitioned into $L$ frames with duration $T_{\textrm{out}}$. Each user $k$ has its dataset $\mathcal{D}_k=\{\boldsymbol{X}_k,\boldsymbol{y}_k\}$, where $\boldsymbol{X}_{k}=[\boldsymbol{x}_{k,1},\boldsymbol{x}_{k,2}, \ldots,\boldsymbol{x}_{k,M_k}]$, and $\boldsymbol{y}_k=[y_{k,1},y_{k,2},\ldots,y_{k,M_k}]$ represent the input and output of user $k$, $M_k$ represents the number of the samples collected by user $k$. Let $M=\sum_{k=1}^{K}M_k$ be the total number of training data samples and $\boldsymbol{w}_k(i)$ be the weight vector of the local FL model of user $k$ at frame $i$. The BS aims to fit the weight vector of the local FL model so as to minimize a particular loss function by using the whole dataset from all the available users. Hence, the global loss function can be expressed as $\frac{1}{K} \sum_{k=1}^{K} \sum_{m=1}^{M_k} f(\boldsymbol{w}_k(i),\boldsymbol{x}_{k,m},{y}_{k,m})$,
where $f(\boldsymbol{w}_k(i),\boldsymbol{x}_{k,m},{y}_{k})$ represents the loss function associated with data point $\boldsymbol{x}_{k,m}$. The important notations are summarized in Table~\ref{tab1}.

After receiving the local FL models of users, the BS will generate the global FL model. Let $\chi_k(i)$ be a Boolean parameter that is set to 1 if user $k$ at frame $i$ is scheduled and to 0 otherwise. Since only a subset of scheduled users that satisfy the minimum signal-to-interference-plus-noise ratio (SINR) $\gamma_{th}$ constraint at frame $i$ will transmit the local FL models to the BS, the update of the global FL model is expressed in a function of the vector of the user scheduling index $\boldsymbol{\chi}(i)=[\chi_1(i),\ldots,\chi_K(i)]$ as follow
\begin{equation}
\label{eq:1}
\boldsymbol{q}(\boldsymbol{\chi}(i))=\frac{\sum_{k=1}^{K} M_k   \chi_k(i) \boldsymbol{w}_{k}(i)}{\sum_{k=1}^{K} M_k \chi_k(i)}.
\end{equation}
Hence, the global loss function is given by
\begin{equation}
\label{eq:2}
F(\boldsymbol{q})=\frac{1}{K} \sum_{k=1}^{K} \sum_{m=1}^{M_k} f(\boldsymbol{q}(\boldsymbol{\chi}(i)),\boldsymbol{x}_{k,m},{y}_{k,m}).
\end{equation}
The gradient descent method is performed to update the local FL model as fellow
\begin{equation}
\label{eq:3}
\boldsymbol{w}_{k}(i)=  \boldsymbol{q}(\boldsymbol{\chi}(i)) -\frac{\lambda}{M_{k}} \sum_{m=1}^{M_{k}} \nabla f\left( \boldsymbol{q}(\boldsymbol{\chi}(i)) , \boldsymbol{x}_{k,m}, {y}_{k,m}\right),
\end{equation}
where $\lambda$ is the learning rate and $\nabla f\left( \boldsymbol{q}(\boldsymbol{\chi}(i)) , \boldsymbol{x}_{k,m},{y}_{k,m}\right)$ is the gradient of $f\left( \boldsymbol{q}(\boldsymbol{\chi}(i)) , \boldsymbol{x}_{k,m}, {y}_{k,m}\right)$ with
respect to $\boldsymbol{q}(\boldsymbol{\chi}(i))$.

\subsection{Channel and Signal Model}
Let $\boldsymbol{h}_k(i) \in \mathbb{C}^{N\times 1}$ be the small-scale fading channel vector for user $k$ at frame $i$, which is assumed to be a quasi-static Gaussian independent and identically distributed (i.i.d.) slow fading channel. Considering only path loss, the large-scale fading component is expressed as $\beta_k=\zeta \frac{d_k^{-\nu}}{d_0^{-\nu}}$, where $\nu$ is the path loss exponent, $d_k$ is the distance between the BS and user $k$, $d_0$ is the reference distance, and $\zeta$ is a constant related to the carrier frequency and reference  distance. Hence, the channel vector $\boldsymbol{g}_k(i) \in \mathbb{C}^{N\times 1}$ for user $k$ at frame $i$ is given by $\boldsymbol{g}_k (i)=\beta^{1/2}_k \boldsymbol{h}_k (i)$ and the channel matrix is defined as $\boldsymbol{G}(i)=[\boldsymbol{g}_1(i), \boldsymbol{g}_2(i),\ldots,\boldsymbol{g}_K(i)] \in \mathbb{C}^{N \times K }$. The BS uses the channel estimation method in~\cite{prec1} for massive MIMO. In particular, the devices send pilot signals with their energy levels to the BS which estimates the channel on the uplink using minimum mean square error (MMSE) estimation. Due to the channel reciprocity in time division duplex (TDD), the channel in the downlink is considered similar to uplink.

Linear receiver techniques such as zero forcing (ZF) are used to achieve near-optimal performance in our system~\cite{prec1}. In particular, ZF has been demonstrated to be a practical beamforming method for massive MIMO systems~\cite{prec1}. Thus, the uplink decoder matrix $\boldsymbol{Z}(i)$ at frame $i$ is given by~$\boldsymbol{Z}(i)= \boldsymbol{G}(i) \boldsymbol{A}(i)$, where $\boldsymbol{A}(i)= (\boldsymbol{G}^H(i) \boldsymbol{G}(i))^{-1}$ and $(.)^H$ represents the Hermitian matrix. The received uplink SINR for user $k$ assuming perfect CSI is expressed as
\begin{equation}
 \label{eq:4}
   \gamma_k(i)=\frac{ p_k(i) }{ [\boldsymbol{A}(i)]_{k,k} \sigma^2 },
\end{equation}
where $p_k(i)$ is the power allocated to user $k$ at frame $i$ and $\sigma^2$ is the noise variance that is assumed to be additive white Gaussian noise (AWGN) with zero mean.

Each user is assumed to be powered by its own energy harvesting source. The harvested energy at each user, which is modeled by a compound Poisson stochastic process~\cite{rami}, is first stored in a battery with maximal capacity $B_{\textrm{max}}$. Let $E_k(i)$ and $B_k(i)$ denote respectively the amount of harvested energy and the battery level of user $k$ at frame $i$. $E_{c}=\varrho  J_k$ is the energy consumption of user $k$ training the local FL model at its own user, where $\varrho$ is an energy consumption coefficient and $J_k$ is the size of the input data of user $k$. The energy consumption for FL model transmission cannot exceed the battery level. Hence, the energy causality constraint is given by
\begin{equation}
 \label{eq:5}
 E_{c} + p_{k}(i) T_{\textrm{out}} \leq B_k(i),
\end{equation}
and the battery level update is expressed as
\begin{equation}
 \label{eq:6}
   B_k(i+1)=\text{min}(B_{\textrm{max}}, B_k(i)-p_{k}(i)  T_{\textrm{out}} + E_k(i)).\\
\end{equation}
Due to limited availability of energy and multiuser interference, only a subset of users can perform FL model update at each time frame. 

\subsection{Problem Formulation}
The aim of this work is to minimize the training loss of the FL model while making use of the available harvested energy and ensuring a minimum SINR to each scheduled user. Hence, the joint energy management and user scheduling problem in FL over an energy harvesting wireless system can be formulated as
\begin{equation}
\label{eq:10}
\begin{split}
\underset{ \underset{k=1,\ldots,K,i=1,\ldots,L} {\{ \chi_k(i),p_k(i)\}}  } {\min}  F(\boldsymbol{q})
\end{split}
\end{equation}
\vspace{-0.3cm}
\begin{align*}\label{c1}
\setlength{\abovedisplayskip}{-20 pt}
\setlength{\belowdisplayskip}{-20 pt}
&\!\!\!\!\!\!\!\!\rm{s.\;t.}\;\scalebox{1}{$\gamma_k(i) \geq \chi_k(i)  \gamma_{th},\;\;\forall k=1,\ldots,K,i=1,\ldots,L,$}\tag{\theequation a}\\
&\scalebox{1}{$\;\;\; \sum\limits_{i=1}^{l} \chi_k(i) (p_k(i)  T_{\textrm{out}} +  E_c) \leq \sum\limits_{i=1}^{l} E_k(i),$}\\
&\scalebox{1}{$\;\;\;\forall k=1,\ldots,K,l=1,\ldots,L,$} \tag{\theequation b}\\
&\scalebox{1}{$\;\;\; \sum\limits_{i=1}^{l} E_k(i) - \sum\limits_{i=1}^{l-1} \chi_k(i) (p_k(i)  T_{\textrm{out}} + E_c) \leq B_{\textrm{max}},$} \\
&\scalebox{1}{$\;\;\;\forall k=1,\ldots,K,l=2,\ldots,L,$} \tag{\theequation c}\\
&\scalebox{1}{$\;\;\; p_k(i) \geq 0,\;\;\forall k=1,\ldots,K,i=1,\ldots,L, $} \tag{\theequation d}\\
&\scalebox{1}{$\;\;\; \chi_k(i) \in \{0,1\},\;\;\forall k=1,\ldots,K,i=1,\ldots,L. $} \tag{\theequation e}\\
\end{align*}
Constraint (\ref{eq:10}a) ensures a minimum received SINR, denoted $\gamma_{th}$, to each user. Constraint (\ref{eq:10}b) is related to the energy causality, i.e. the consumed energy at user $k$ cannot exceed the available energy at the battery. Additionally, constraint (\ref{eq:10}c) implies that the harvested energy at the current frame cannot exceed the maximal battery capacity. Finally, constraint (\ref{eq:10}d) ensures the non-negativity of the allocated amounts of power.\\
The formulated problem $(\ref{eq:10})$ is a mixed-integer non-linear problem (MINLP) because of its combinatorial nature and the non-linearity of the objective function. Meanwhile, to solve problem $(\ref{eq:10})$, the impact of the transmit power $p_k(i)$ and the user scheduling $\chi_k(i)$ on the training loss has to be analyzed.

\section{Impact of Wireless Parameters on FL Performance}
To solve (\ref{eq:10}), we first find the relationship between the user scheduling and the FL performance. Hence, the convergence rate of FL is analyzed as following.\\ 
\textbf{Theorem 1.} The expected convergence rate of the FL algorithm is given by
\begin{equation}
\label{eq:11}
\begin{aligned}
\mathbb{E}\left[F\left(\boldsymbol{q}_{i+1}\right)-F\left(\boldsymbol{q}^{*}\right)\right]  &\leq \sum_{t=0}^{i} \left(\prod_{j=t+1}^{i} b_j\right) a_t  \\
& + \mathbb{E}\left(F\left(\boldsymbol{q}_{0}\right)-F\left(\boldsymbol{q}^{*}\right)\right) \prod_{t=0}^{i} b_t,
\end{aligned}
\end{equation}
where
\begin{equation}
\label{eq:12}
a_t=\frac{2 \zeta_{1}}{V M} \sum_{k=1}^{K} M_{k}\left(1-\chi_k(t)\right),
\end{equation}
\begin{equation}
\label{eq:13}
b_t=1-\frac{\mu}{V}+\frac{4 \mu \zeta_{2}}{V M} \sum_{k=1}^{K} M_{k}\left(1-\chi_k(t)\right),
\end{equation}
$\boldsymbol{q}^{*}$ is the optimal FL model that the FL targets to converge to, and $F\left(\boldsymbol{q}^{*}\right)=\frac{1}{K} \sum_{k=1}^{K} \sum_{m=1}^{M_k} f(\boldsymbol{q}^{*},\boldsymbol{x}_{k,m},{y}_{k,m})$ is the optimal training loss, $V$ is a Lipschitz coefficient, and the coefficients $\zeta_{1},\zeta_{2}$, and $\mu$ are given by the second-order Taylor expansion of the loss function which is assumed to be convex.
\begin{proof}
See in Appendix A.\\
\end{proof}
From $(\ref{eq:11})$, it can be seen that a gap $\sum_{t=0}^{i-1} \left(\prod_{j=t+1}^{i-1} b_j\right) a_t$ exists between $\mathbb{E}\left[F\left(\boldsymbol{q}_{i}\right)\right]$ and $\mathbb{E}\left[F\left(\boldsymbol{q}^*\right)\right] $. This gap is caused by user scheduling policy which decreases when the number of scheduled users increases. Hence, maximizing the number of scheduled users allows to enhance the convergence speed of the FL algorithm.\\
Considering that all users are scheduled $\chi_k(i)=1$, we have $a_i=0$ and $b_i=1-\frac{2 \mu}{V}$. Hence, the expected convergence rate of the FL algorithm is given by
\begin{equation}
\label{eq:20}
\mathbb{E}\left[F\left(\boldsymbol{q}_{i+1}\right)-F\left(\boldsymbol{q}^{*}\right)\right]  \leq  \mathbb{E}\left(F\left(\boldsymbol{q}_{0}\right)-F\left(\boldsymbol{q}^{*}\right)\right)  \left(1-\frac{2 \mu}{V}\right)^i.
\end{equation}
The relationship between the FL convergence rate and the wireless parameters could be used to simplify and to solve the main problem (\ref{eq:10}).

\section{Optimal Resource Allocation}
\subsection{Algorithm Design}
After establishing the relationship between the FL convergence rate and wireless parameters,  the objective function of the main problem (\ref{eq:10}) can be simplified to $\sum\limits_{t=0}^{i} \left(\prod_{j=t+1}^{i} b_j\right) a_t$ based on the inequality $(\ref{eq:11})$. The main problem $(\ref{eq:10})$ can be simplified and decomposed into sub-problems. The objective function at frame $i$ can be expressed using $(\ref{eq:16})$ as $a_i+b_i \mathbb{E}\left(F\left(\boldsymbol{q}_{i}\right)-F\left(\boldsymbol{q}^{*}\right)\right)$. Hence, the sub-problem at frame $i$ is given by
\begin{equation}
\label{eq:31}
\begin{split}
\underset{ \underset{k=1,\ldots,K} {\{p_k(i), \chi_k(i)\}}  } {\min} a_i+b_i \mathbb{E}\left(F\left(\boldsymbol{q}_{i}\right)-F\left(\boldsymbol{q}^{*}\right)\right)
\end{split}
\end{equation}
\vspace{-0.3cm}
\begin{align*}\label{c1}
\setlength{\abovedisplayskip}{-20 pt}
\setlength{\belowdisplayskip}{-20 pt}
&\!\!\!\!\!\!\rm{s.\;t.}\;\scalebox{1}{$\gamma_k(i) \geq \chi_k(i) \gamma_{th},\;\;\forall k=1,\ldots,K,$} \tag{\theequation a}\\
&\scalebox{1}{$\;\;\;\chi_k(i)  (p_k(i) T_{\textrm{out}} +  E_c) \leq  B_k(i),\;\;\forall k=1,\ldots,K,$} \tag{\theequation b}\\
&\scalebox{1}{$\;\;\; B_k(i) + E_k(i) -  \chi_k(i)  ( p_k(i) T_{\textrm{out}} +  E_c) \leq B_{\textrm{max}},$} \\
&\scalebox{1}{$\;\;\;\forall k=1,\ldots,K,$} \tag{\theequation c}\\
&\scalebox{1}{$\;\;\; p_k(i) \geq 0,\;\;\forall k=1,\ldots,K,$} \tag{\theequation d}\\
&\scalebox{1}{$\;\;\; \chi_k(i) \in \{0,1\},\;\;\forall k=1,\ldots,K.$} \tag{\theequation e}\\
\vspace{-0.2cm}
\end{align*}
The impact of energy consumption can be investigated. The required energy to train the desired FL model for a certain number of communication rounds is derived. The required transmit power to meet the SINR constraint of user $k$ at frame $i$ is given by
\begin{equation}
\label{eq:21}
p_k(i) = \chi_k(i) \gamma_{th} [\boldsymbol{A}(i)]_{k,k} \sigma^2.
\end{equation}
Hence, the required energy for a successful transmission of the weight vector $\boldsymbol{w}_k(i)$ of user $k$ at frame $i$ is given by
\begin{equation}
\label{eq:22}
e_k(i) = \chi_k(i) \left( \gamma_{th} [\boldsymbol{A}(i)]_{k,k} \sigma^2 + E_c \right).
\end{equation}
Hence, the problem (\ref{eq:31}) can be simplified as follows
\begin{equation}
\label{eq:3111}
\begin{split}
\underset{ \underset{k=1,\ldots,K} {\{ \chi_k(i)\}}  } {\min} a_i+b_i \mathbb{E}\left(F\left(\boldsymbol{q}_{i}\right)-F\left(\boldsymbol{q}^{*}\right)\right)
\end{split}
\end{equation}
\vspace{-0.3cm}
\begin{align*}\label{c1}
\setlength{\abovedisplayskip}{-20 pt}
\setlength{\belowdisplayskip}{-20 pt}
&\!\rm{s.\;t.}\;\scalebox{1}{$\chi_k(i)  (\gamma_{th} [\boldsymbol{A}(i)]_{k,k} \sigma^2 T_{\textrm{out}} +  E_c) \leq  B_k(i),$}\\
&\scalebox{1}{$\;\;\; \forall k=1,\ldots,K,$} \tag{\theequation a}\\
&\scalebox{1}{$\;\;\; B_k(i) + E_k(i) -  \chi_k(i)  (\gamma_{th} [\boldsymbol{A}(i)]_{k,k} \sigma^2  T_{\textrm{out}} +  E_c) \leq B_{\textrm{max}},$}\\
&\scalebox{1}{$\;\;\;\forall k=1,\ldots,K,$} \tag{\theequation b}\\
&\scalebox{1}{$\;\;\; \chi_k(i) \in \{0,1\},\;\;\forall k=1,\ldots,K.$} \tag{\theequation c}\\
\end{align*}
The objective function in $(\ref{eq:3111})$ can be rewritten by
\begin{equation}
\label{eq:311}
\begin{aligned}
a_i+b_i u_i &  = \frac{2 \zeta_{1}}{V M} \sum_{k=1}^{K} M_{k}\left(1-\chi_k(i)\right) \\
& + \left(1-\frac{\mu}{V}+\frac{4 \mu \zeta_{2}}{V M} \sum_{k=1}^{K} M_{k}\left(1-\chi_k(i)\right)\right) u_i\\
&  = \left( \frac{2 \zeta_{1}}{V M} + \frac{4 \mu \zeta_{2}}{V M} u_i \right)\sum_{k=1}^{K}  M_{k}\left(1-\chi_k(i)\right) \\ 
& +  \left(1-\frac{\mu}{V}\right) u_i.
\end{aligned}
\end{equation}
Hence, the objective function could be simplified to the equivalent function $-\sum_{k=1}^{K}  M_{k} \chi_k(i)$. Meanwhile, the constraints could be simplified and the sub-problem is reformulated as
\vspace{-0.2cm}
\begin{equation}
\label{eq:32}
\begin{split}
\underset{ \underset{k=1,\ldots,K} {\{ \chi_k(i)\}}  } {\min} -\sum_{k=1}^{K}  M_{k} \chi_k(i)
\end{split}
\end{equation}
\vspace{-0.3cm}
\begin{align*}\label{c1}
\setlength{\abovedisplayskip}{-20 pt}
\setlength{\belowdisplayskip}{-20 pt}
&\!\!\!\!\!\!\!\!\rm{s.\;t.}\;\scalebox{1}{$\chi_k(i)  \leq  \frac{B_k(i)}{\gamma_{th} [\boldsymbol{A}(i)]_{k,k} \sigma^2  T_{\textrm{out}} +  E_c},\;\;\forall k=1,\ldots,K,$}\tag{\theequation a}\\
&\scalebox{1}{$\;\;\; -  \chi_k(i)  \leq \frac{B_{\textrm{max}}-B_k(i)-E_k(i)}{\gamma_{th} [\boldsymbol{A}(i)]_{k,k} \sigma^2  T_{\textrm{out}} +  E_c},\;\;\forall k=1,\ldots,K,$} \tag{\theequation b}\\
&\scalebox{1}{$\;\;\; \chi_k(i) \in \{0,1\},\;\;\forall k=1,\ldots,K.$} \tag{\theequation c}\\
\end{align*}

Solving $(\ref{eq:32})$ is equivalent to maximize the number of scheduled users satisfying the constraints (\ref{eq:32}a)-(\ref{eq:32}c). Hence, the optimal user scheduling is given by
\begin{equation}
\label{eq:33}
\chi_k(i)=\left\{\begin{array}{l}
1, B_k(i)+E_k(i)-B_{\textrm{max}} \leq \gamma_{th} [\boldsymbol{A}(i)]_{k,k} \sigma^2  T_{\textrm{out}} \\
~~~~~~~~~~~~~~~~~~~~~~~~~~~~~~~~+  E_c \leq B_k(i), \\
0, \quad \text { otherwise}.
\end{array}\right.
\end{equation}

From (\ref{eq:33}), we can observe that user scheduling strongly depends on the channel coefficients, available energy, and the maximal battery capacity. The proposed user scheduling and energy management algorithm (USEM) is summarized in \textbf{Algorithm~1}.

\begin{algorithm}[t]
\begin{algorithmic}
\caption{Optimal User Scheduling and Energy Management Algorithm (USEM)}
\State 
\State $B_k(1) \gets E_k(1),~k=1..K$, // initialization
\For{$i=1:L$}
	\For{$k=1:K$} 
	     \If{$B_k(i)+E_k(i)-B_{\textrm{max}} \leq \gamma_{th} [\boldsymbol{A}(i)]_{k,k} \sigma^2 $}
 	        \State  $\chi_k(i) \gets 1$
 	        \State $p_k(i) \gets \gamma_{th} [\mathbf{A}(i)]_{k,k} \sigma^2$, power allocation
 	        \State $e_k(i) \gets p_k(i) T_{\textrm{out}}+ E_{c}$, compute the required power
 	        \State Send the weight vector $\boldsymbol{w}_k(i)$ to the BS
 	     \Else
 	        \State $\chi_k(i) \gets 0$
 	        \State $e_k(i) \gets 0$
	     \EndIf 
	     \If{ $i < L$}
	         \State $B_k(i+1) \gets \min \left(B_{\textrm{max}}, B_k(i)-e_k(i)+E_k(i+1)\right)$, batteries update
	     \EndIf	       		
	\EndFor
	\State $\boldsymbol{q} \gets \frac{\sum_{k=1}^{K} M_k   \chi_k(i) \boldsymbol{w}_{k}(i)}{\sum_{k=1}^{K} M_k \chi_k(i)}$, update the global model 
\EndFor
\end{algorithmic}
\end{algorithm}

The proposed algorithm can be easily extended to users with multiple antennas. In particular, ZF beamforming may be changed to block diagonalization (BD) beamforming. Indeed, when the users are equipped with multiple antennas, the multiuser interference can be suppressed using low complexity BD, which is known to achieve near-optimal capacity~\cite{bd,bd2}.

\subsection{Implementation and Complexity}
In this section, the implementation and the complexity of the algorithm USEM are analyzed. To implement USEM, the BS must first calculate the two terms $B_k(i)+E_k(i)-B_{\textrm{max}}$ and $\gamma_{th} [\boldsymbol{A}(i)]_{k,k} \sigma^2$ for each user. To calculate the first term, the BS must know the battery level $B_k(i)$ and the amount of harvested energy $E_k(i)$ of each user. The battery level $B_k(i)$ at frame $i$ can be computed once the BS calculates the consumed energy $e_k(i-1)$ and knows the amount of harvested energy $E_k(i-1)$ at frame $i-1$ of each user. The amount of harvested energy could be sent directly by the users to the BS or it can be estimated by the BS using energy prediction tools~\cite{rami2}. Next, to calculate the second term, the BS must calculate the matrix $\boldsymbol{A}(i)$ which depend on the channel matrix $\boldsymbol{G}(i)$. The BS can use channel estimation techniques for massive MIMO such as minimum mean square error~\cite{prec1} to estimate $\boldsymbol{G}(i)$. After calculating these two terms, the BS performs user scheduling and update the batteries levels by calculating the consumed energy $e_k(i)$. Next, the FL mechanism is performed, the BS must send the FL model information to the scheduled users users and receive the local updates. Finally, the BS update the global model.

The complexity of USEM could be evaluated. The matrix $\boldsymbol{A}(i)$ is calculated using one matrix multiplication with a complexity order of $O(N K^2)$ and one matrix inversion with a complexity order of $O(K^3)$. The others terms such as the consumed energy are calculated with simple operations. Hence, the user scheduling for the $L$ frames is performed with a polynomial complexity $O\left(L(N K^2+K^3)\right)$.

\section{Multiple Base Stations}
In this section, the system model is extended from one BS to multiple BSs in order to implement FL over a large scale network which consists of more BSs and more users. For multiple BSs, we need to consider the interference between different BSs as well as different users. Moreover, we need to optimize the user association with different BSs while considering energy and QoS constraints. The association decision is based on large-scale fading. The user association index is defined as
\begin{equation}
\label{eq:40}
\alpha_{k,s}=\left\{\begin{array}{ll}1, & \text { if user } k \text { is associated to } \mathrm{BS}~s, \\ 0, & \text {otherwise}.
\end{array}\right.
\end{equation}
Let $S$ denotes the number of BSs. Each user is assumed to be associated to at most one BS as
\begin{equation}
\label{eq:41}
\sum_{s=1}^{S} \alpha_{k,s} \leq 1,\;\;\forall k=1,\ldots,K.
\end{equation}

Each BS is assumed to have perfect channel state information of all users in the network. Hence, each BS may apply a cooperative ZF scheme to remove the intra-cell and inter-cell interference as in~\cite{zf,zf2}. The ZF decoder matrix $\boldsymbol{A}_{s}(i)$ of BS $s$ at frame $i$ is given by $\boldsymbol{A}_{s}(i)=\boldsymbol{G}_{s}(i)\left(\boldsymbol{G}_{s}(i)^{H} \boldsymbol{G}_{s}(i)\right)^{-1}$. The SINR between user $k$ and BS $s$ at frame $i$ can be written as
\begin{equation}
\label{eq:42}
\gamma_{k,s}(i)=\frac{ p_{k,s}(i) }{[\boldsymbol{A}_{s}(i)]_{k,k} \sigma^2 },
\end{equation}
where $p_{k,s}(i)$ is the transmit power from user $k$ to BS $s$.

Given a certain user association configuration, each BS can schedule users for local FL model transmissions. Hence, the joint user association, user scheduling, and power allocation problem in FL is formulated as (\ref{eq:43}).\\
From constraints (\ref{eq:43}b) and (\ref{eq:43}c), it can be seen that when $\alpha_{k,s}=0$, the user scheduling index  $\chi_{k,s}(i)$ is forced to 0. However, when $\alpha_{k,s}=1$ the user scheduling index  $\chi_{k,s}(i)$ can be either 0 or 1.

\begin{equation}
\label{eq:43}
\begin{split}
\underset{ \underset{k=1,\ldots,K,i=1,\ldots,L,s=1,\ldots,S} {\{ \alpha_{k,s},\chi_{k,s}(i),p_{k}(i)\}}  } {\min} F(\boldsymbol{q})
\end{split}
\end{equation}
\vspace{-0.3cm}
\begin{align*}\label{c1}
\setlength{\abovedisplayskip}{-20 pt}
\setlength{\belowdisplayskip}{-20 pt}
&\!\!\!\!\rm{s.\;t.}\;\scalebox{1}{$\gamma_{k,s}(i) \geq \alpha_{k,s} \chi_{k,s}(i) \gamma_{th},\;\;\forall k=1,\ldots,K,$}\\
&\scalebox{1}{$\;\;\; s=1,\ldots,S,i=1,\ldots,L,$}\tag{\theequation a}\\
&\scalebox{1}{$\;\;\; \sum_{s=1}^{S} \alpha_{k,s} \leq 1,\;\;\forall k=1,\ldots,K,$} \tag{\theequation b}\\
&\scalebox{1}{$\;\;\;  \sum_{i=1}^{l} \chi_{k,s}(i) (p_{k,s}(i)  T_{\textrm{out}} +  E_c) \leq \alpha_{k,s} \sum_{i=1}^{l} E_k(i),$}\\
&\scalebox{1}{$\;\;\;\forall k=1,\ldots,K,l=1,\ldots,L,s=1,\ldots,S,$} \tag{\theequation c}\\
&\scalebox{1}{$\;\;\;  - \sum_{i=1}^{l-1} \chi_{k,s}(i) (p_{k,s}(i) T_{\textrm{out}} + E_c) \leq \alpha_{k,s}  (B_{\textrm{max}}-$} \\
&\scalebox{1}{$\;\;\; \sum_{i=1}^{l} E_k(i)), \forall k=1,\ldots,K,l=2,\ldots,L,s=1,\ldots,S,$} \tag{\theequation d}\\
&\scalebox{1}{$\;\;\;  p_{k,s}(i) \geq 0,\;\;\forall k=1,\ldots,K,i=1,\ldots,L,s=1,\ldots,S,$} \tag{\theequation e}\\
&\scalebox{1}{$\;\;\; \alpha_{k,s} \in \{0,1\},\;\;\forall k=1,\ldots,K,s=1,\ldots,S. $} \tag{\theequation f}\\
&\scalebox{1}{$\;\;\; \chi_{k,s}(i) \in \{0,1\},\;\;\forall k=1,\ldots,K,i=1,\ldots,L,$}\\ 
&\scalebox{1}{$\;\;\; s=1,\ldots,S. $} \tag{\theequation g}
\end{align*}

The inequality in $(\ref{eq:11})$ that rely on the FL model and wireless parameters is adapted to the multi-BS system model by considering the user association index $\sum_{s=1}^{S} \alpha_{k,s}$, where $\sum_{s=1}^{S} \alpha_{k,s}=1$ implies that user $k$ is associated with one BS, otherwise, we have $\sum_{s=1}^{S} \alpha_{k,s}=0$. The expected convergence rate of the FL algorithm is given by
\begin{equation}
\label{eq:44}
\begin{aligned}
\mathbb{E}\left[F\left(\boldsymbol{q}_{i+1}\right)-F\left(\boldsymbol{q}^{*}\right)\right]  &\leq \sum_{t=0}^{i} \left(\prod_{j=t+1}^{i} d_j\right) c_t \\ 
&+\mathbb{E}\left(F\left(\boldsymbol{q}_{0}\right)-F\left(\boldsymbol{q}^{*}\right)\right) \prod_{t=0}^{i} d_t
\end{aligned}
\end{equation}
where
\begin{equation}
\label{eq:45}
c_t=\frac{2 \zeta_{1}}{V M} \sum_{k=1}^{K} M_{k}\left(1-\sum_{s=1}^{S} \chi_{k,s}(t)\right),
\end{equation}
and
\begin{equation}
\label{eq:46}
d_t=1-\frac{\mu}{V}+\frac{4 \mu \zeta_{2}}{V M} \sum_{k=1}^{K} M_{k}\left(1-\sum_{s=1}^{S} \chi_{k,s}(t) \right).
\end{equation}

The objective function can be simplified based on the inequality $(\ref{eq:44})$ to $\sum\limits_{i=1}^{L} \left(\prod\limits_{j=i+1}^{L} d_j\right) c_i $. The minimization of this term could be also simplified using $(\ref{eq:45})$ and $(\ref{eq:46})$ to minimize\\
$\sum\limits_{i=1}^{L} \sum\limits_{k=1}^{K} M_{k}\left(1-\sum\limits_{s=1}^{S} \chi_{k,s}(i)\right)$. Moreover, the constraints are simplified using $(\ref{eq:22})$. The optimization problem in $(\ref{eq:43})$ can be simplified as (\ref{eq:47}). Hence, the user association and scheduling problem is formulated as an integer linear program which can be solved using branch and bound approaches~\cite{bb}.

\begin{equation}
\label{eq:47}
\begin{split}
\underset{ \underset{k=1,\ldots,K,i=1,\ldots,L,s=1,\ldots,S} {\{ \alpha_{k,s},\chi_{k,s}(i)\}}  } {\min}
\sum\limits_{i=1}^{L} \sum\limits_{k=1}^{K} M_{k}\left(1- \sum\limits_{s=1}^{S} \chi_{k,s}(i)\right)
\end{split}
\end{equation}
\vspace{-0.3cm}
\begin{align*}\label{c1}
\setlength{\abovedisplayskip}{-20 pt}
\setlength{\belowdisplayskip}{-20 pt}
&\!\!\rm{s.\;t.}\;\scalebox{1}{$\sum_{s=1}^{S} \alpha_{k,s} \leq 1,\;\;\forall k=1,\ldots,K,$}\tag{\theequation a}\\
&\scalebox{1}{$\;\;\;  \sum_{i=1}^{l} \chi_{k,s}(i) (\gamma_{th} [\boldsymbol{A}_s(i)]_{k,k} \sigma^2  T_{\textrm{out}} +  E_c) \leq \alpha_{k,s} $}\\
&\scalebox{1}{$\;\;\; \sum_{i=1}^{l} E_k(i), \forall k=1,\ldots,K,l=1,\ldots,L,s=1,\ldots,S,$} \tag{\theequation b}\\
&\scalebox{1}{$\;\;\;  -  \sum_{i=1}^{l-1} \chi_{k,s}(i) (\gamma_{th} [\boldsymbol{A}_s(i)]_{k,k} \sigma^2 T_{\textrm{out}} + E_c) \leq \alpha_{k,s} (B_{\textrm{max}} $} \\
&\scalebox{1}{$\;\;\; -\sum_{i=1}^{l} E_k(i)), \forall k=1,\ldots,K,l=2,\ldots,L,s=1,\ldots,S,$} \tag{\theequation c}\\
&\scalebox{1}{$\;\;\; \alpha_{k,s} \in \{0,1\},\;\;\forall k=1,\ldots,K,s=1,\ldots,S.$} \tag{\theequation d}\\
&\scalebox{1}{$\;\;\; \chi_{k,s}(i) \in \{0,1\},\;\;\forall k=1,\ldots,K,i=1,\ldots,L,$}\\ 
&\scalebox{1}{$\;\;\; s=1,\ldots,S. $} \tag{\theequation e}
\end{align*}
For the user association problem, the calculation of similar parameters to USEM algorithm are required such as matrix inversion and energy computation. The optimization problem has $2 K L S+K$ constraints and $K S$ variables. The optimal user association can be solved using branch and bound technique which may performs in polynomial time~\cite{bb2}. The BS may have sufficient computational resources to implement it.

\section{Numerical Results}
In this section, the performance of the FL over a wireless system powered by energy harvesting is evaluated through experimental simulations. The users are assumed to be randomly distributed within a circular cell. Two scenarios are investigated for the simulations, single BS (scenario 1) and multiple BSs (scenario 2). The simulation parameters used in this section are summarized in Table~\ref{tab2}.
\begin{table}[t]
	\centering
	\caption{Simulation Parameters.}
	\label{tab2}
	\begin{tabular}{|c|c|c|}
		\hline  \textbf{Symbol} & \textbf{Description} &  \textbf{Value}  \\
		\hline   $\nu$ & path loss exponent & 3.7 \\
		\hline   $B_{\textrm{max}}$ & max battery capacity & 50 J \\
		\hline   $N$ & number of antennas & 128 \\
		\hline   $K$ & number of users & 10 \\
		\hline   $L$ & number of frames & 300 \\
		\hline    & noise PSD & -174 dBm/Hz \\
		\hline    & circuit power per RF chain  & 30 dBm\\
		\hline    & cell radius & 500 m \\
		\hline
	\end{tabular}
\end{table}

\begin{figure}[h]
	\centerline{\includegraphics [width=1.05\columnwidth]  {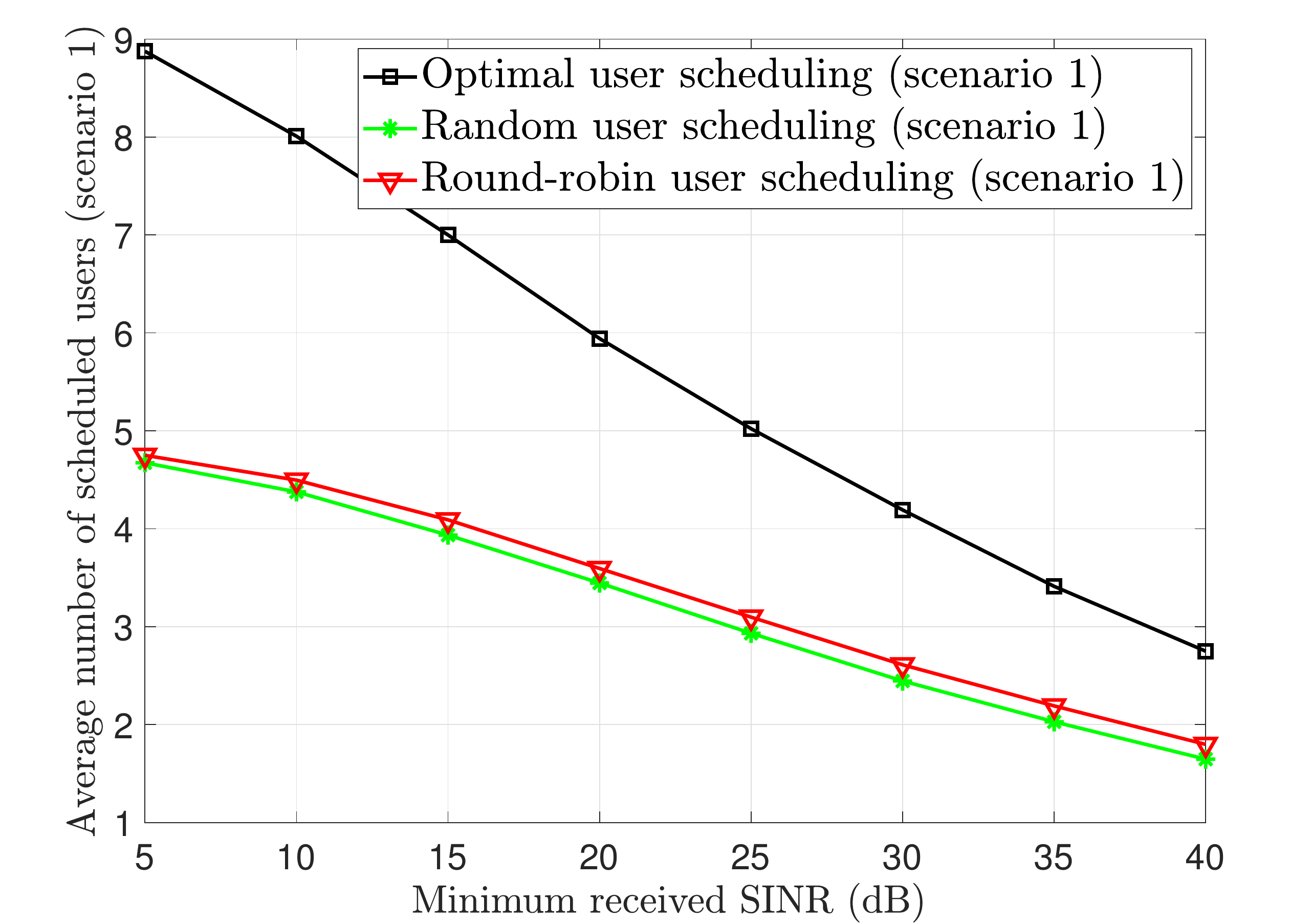}}
	\caption{Average number of scheduled users as a function of the SINR with different scheduling schemes.}
	\label{fig2}
\end{figure}

\begin{figure}[h]
	\centerline{\includegraphics [width=1.05\columnwidth]  {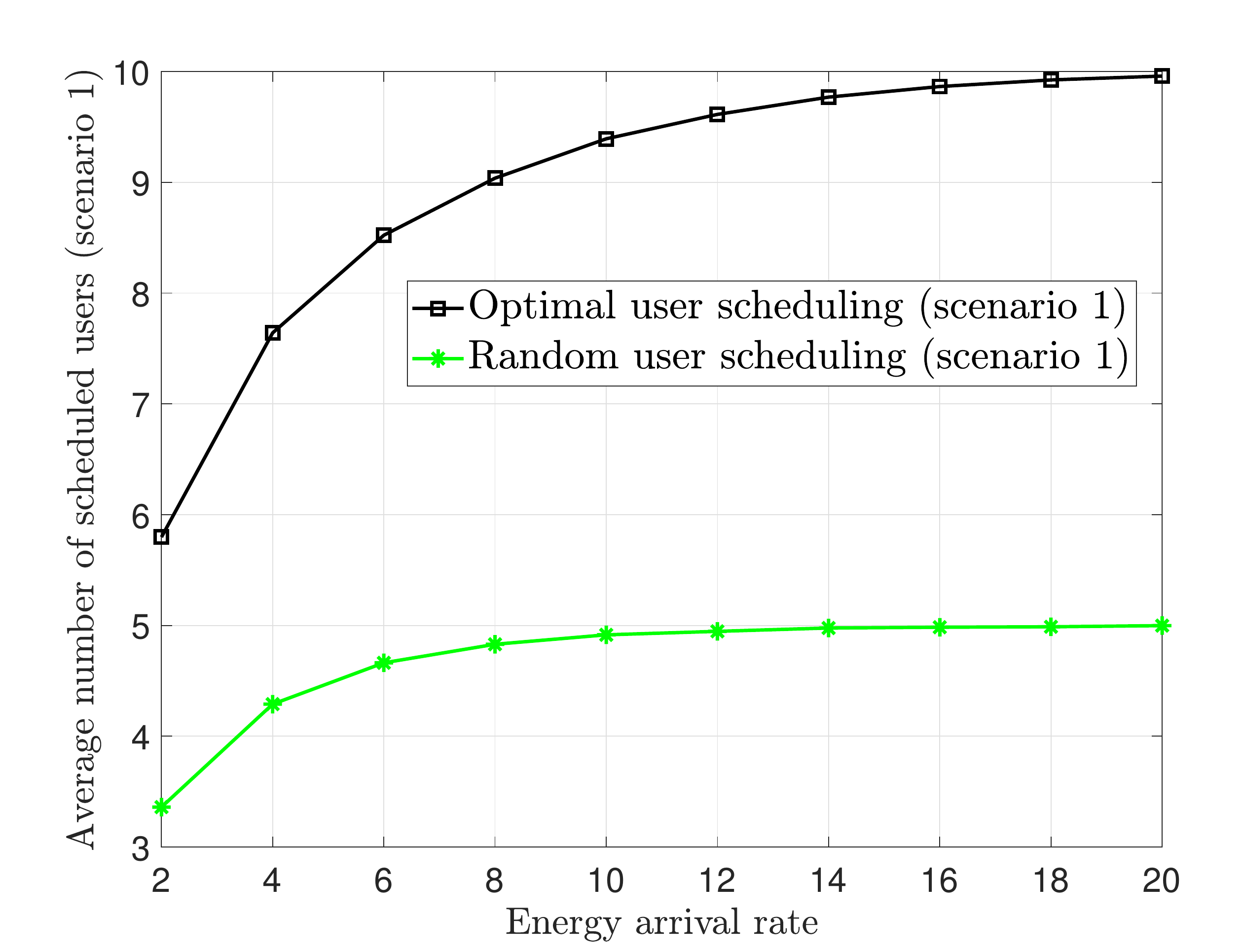}}
	\caption{Average number of scheduled users as a function of the energy arrival rate $\theta$ with different user scheduling schemes considering a SINR threshold  $\gamma_{th}=10$ dB.}
	\label{fig2e}
\end{figure}

In Fig.~\ref{fig2}, the performance of the proposed optimal user scheduling algorithm is investigated by showing the average number of scheduled users versus the minimum received SINR. As expected, the average number of scheduled users is lower when the QoS constraints are more stringent. It can be seen in Fig.~\ref{fig2} that the proposed user scheduling algorithm can increase the number of devices that participate in FL by up to 100\% at 5 dB and 50\% at 40 dB compared to the round robin algorithm. The performance gap between decreases for high SINR threshold because the impact of interference is major. The enhanced performance is due to the efficient management of the harvested energy and deactivating the users that require a large amount of energy.
Fig.~\ref{fig2e} shows the performance of the user scheduling algorithms as a function of the energy arrival rate. It is can be seen that increasing the energy arrival rate allows to increase the number of scheduled devices for both schemes. However, the performance tends to saturate specifically for random scheduling.\\ 

\begin{figure}[h]
	\centerline{\includegraphics [width=1.05\columnwidth]  {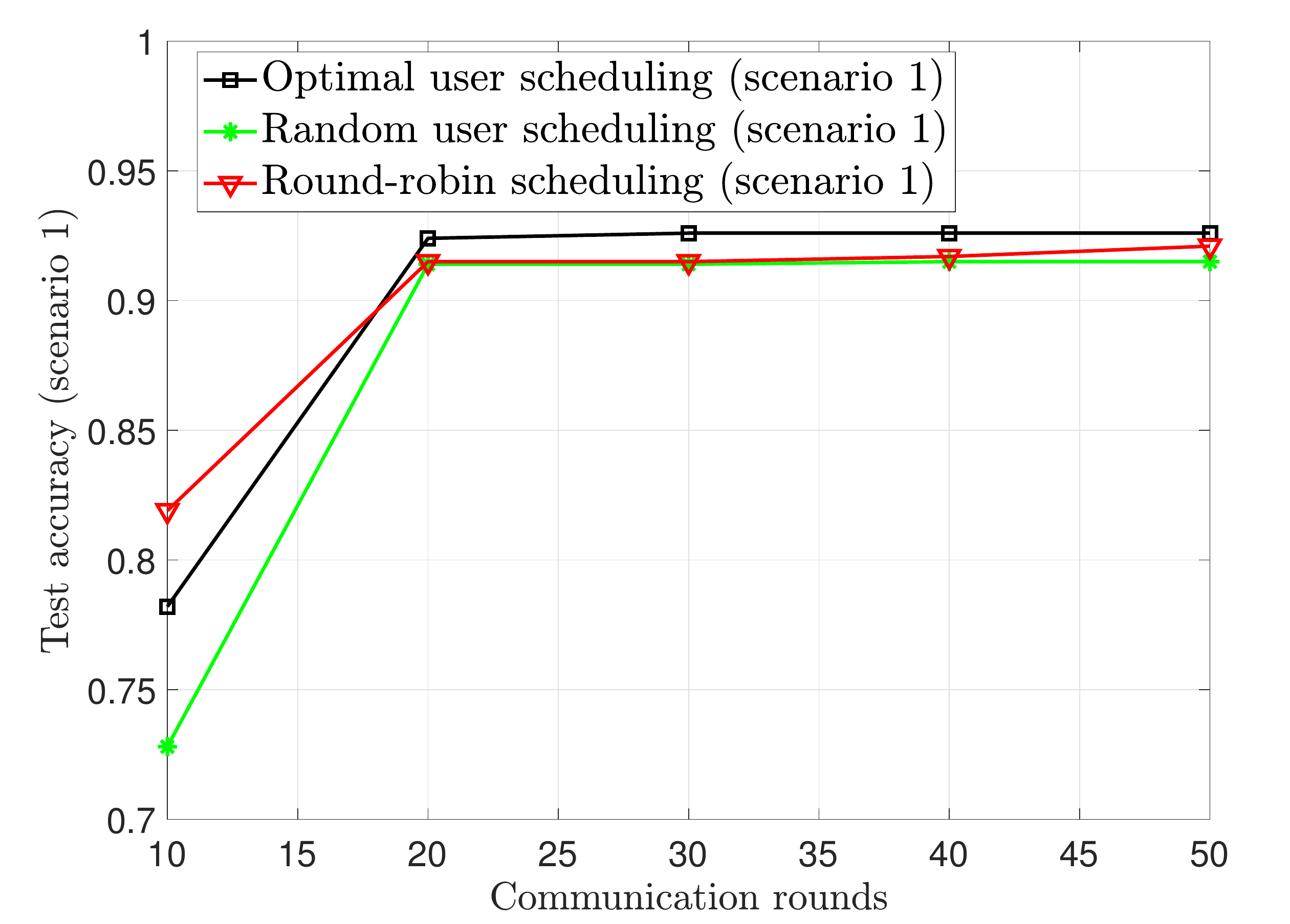}}
	\caption{Test accuracy of the trained SVM with different scheduling schemes considering a SINR threshold  $\gamma_{th}=20$ dB.}
	\label{fig3}
\end{figure}

\begin{figure}[h]
		\centerline{\includegraphics [width=1.05\columnwidth]  {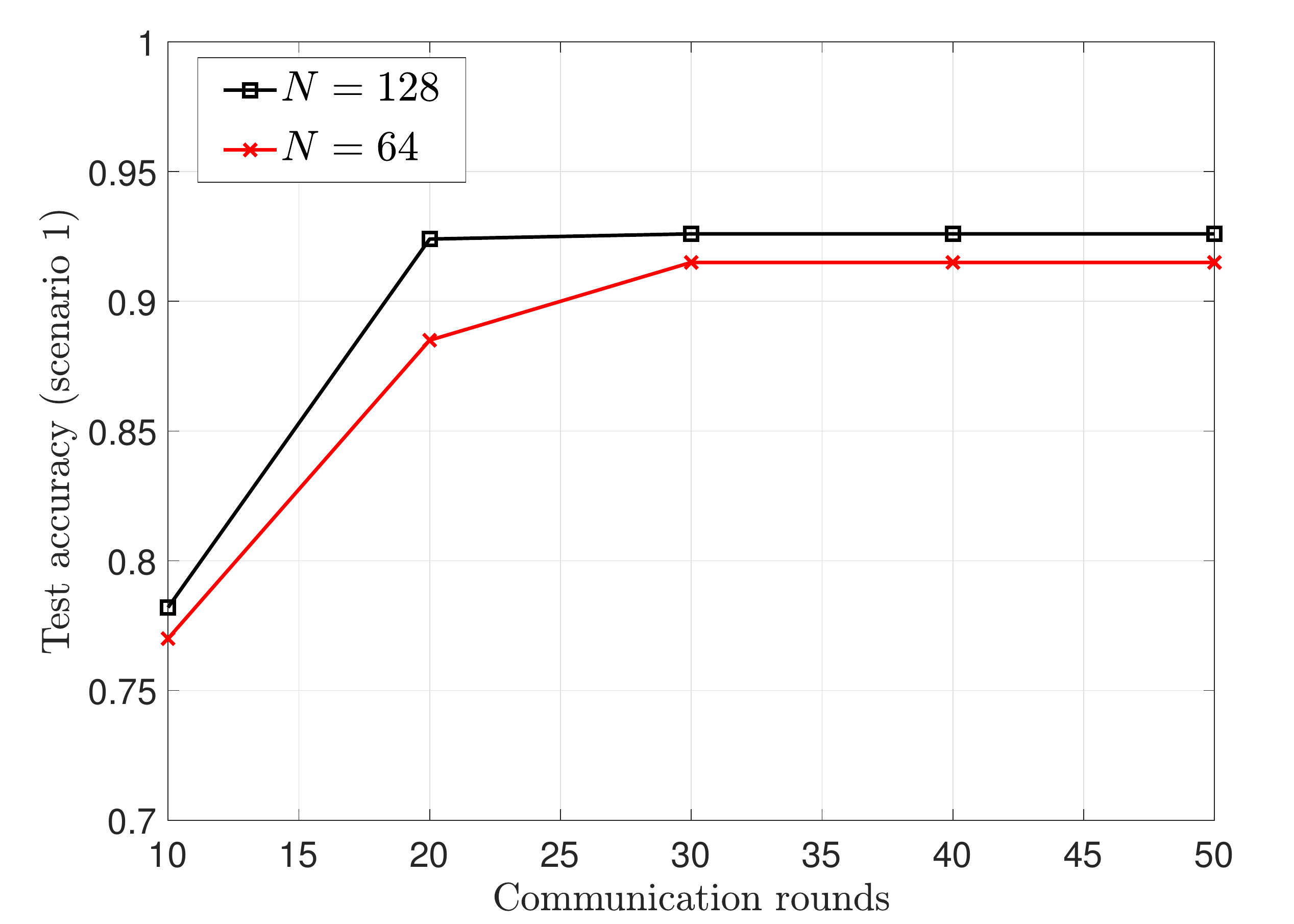}}
		\caption{Impact of the number of antennas $N$ on test accuracy of the trained SVM with optimal user scheduling considering a SINR threshold  $\gamma_{th}=20$ dB.}
		\label{fig3n}
\end{figure}

In Fig.~\ref{fig3}, the effect of scheduling policies on FL performance is shown. The performance of the proposed USEM algorithm is evaluated in FL over wireless environment considering an ML model known as support vector machine (SVM) and a SINR threshold  $\gamma_{th}=20$ dB. A two-class classification task to recognize digits 0 and 8 for each UE with 2 training samples from the MNIST dataset, is considered. This ML model is tested every 10 communications rounds over 100 test samples with a learning rate 0.00001. The results are averaged over five Monte Carlo simulations. A successful global aggregation is defined as when a set of users successfully transmit their local wights to the BS by meeting the SINR and energy requirements and the BS is able to aggregate the received wights. It is clear that the optimal user scheduling algorithm enhances the FL performance in terms of accuracy compared to the random and round-robin scheduling, since it increases the successful global aggregations. This mainly is due to fact that optimal scheduling algorithm can tackle more efficiently the interference and energy constraints. The proposed FL achieving only 93\% accuracy instead of 100\% accuracy implies that the proposed FL cannot correctly identify all the handwritten digits. The identification errors are caused by the FL model, FL training method, and FL training datasets, which are not optimized by the proposed FL. Our purpose is to minimize the training loss instead of reducing training loss to 0. Therefore, the accuracy may be smaller than 100\%. In fact, 100\% accuracy is rarely achieved by any ML model~\cite{us1,us2,us3,us4,us5}.
Fig.~\ref{fig3n} shows the impact of the number of antennas at the BS on FL performance in training SVM. Clearly, an increase in the number of antennas allows an increase in the number of successful global aggregations which significantly improves the test accuracy of the trained model.
\begin{figure}[h]
	\centerline{\includegraphics [width=1.05\columnwidth]  {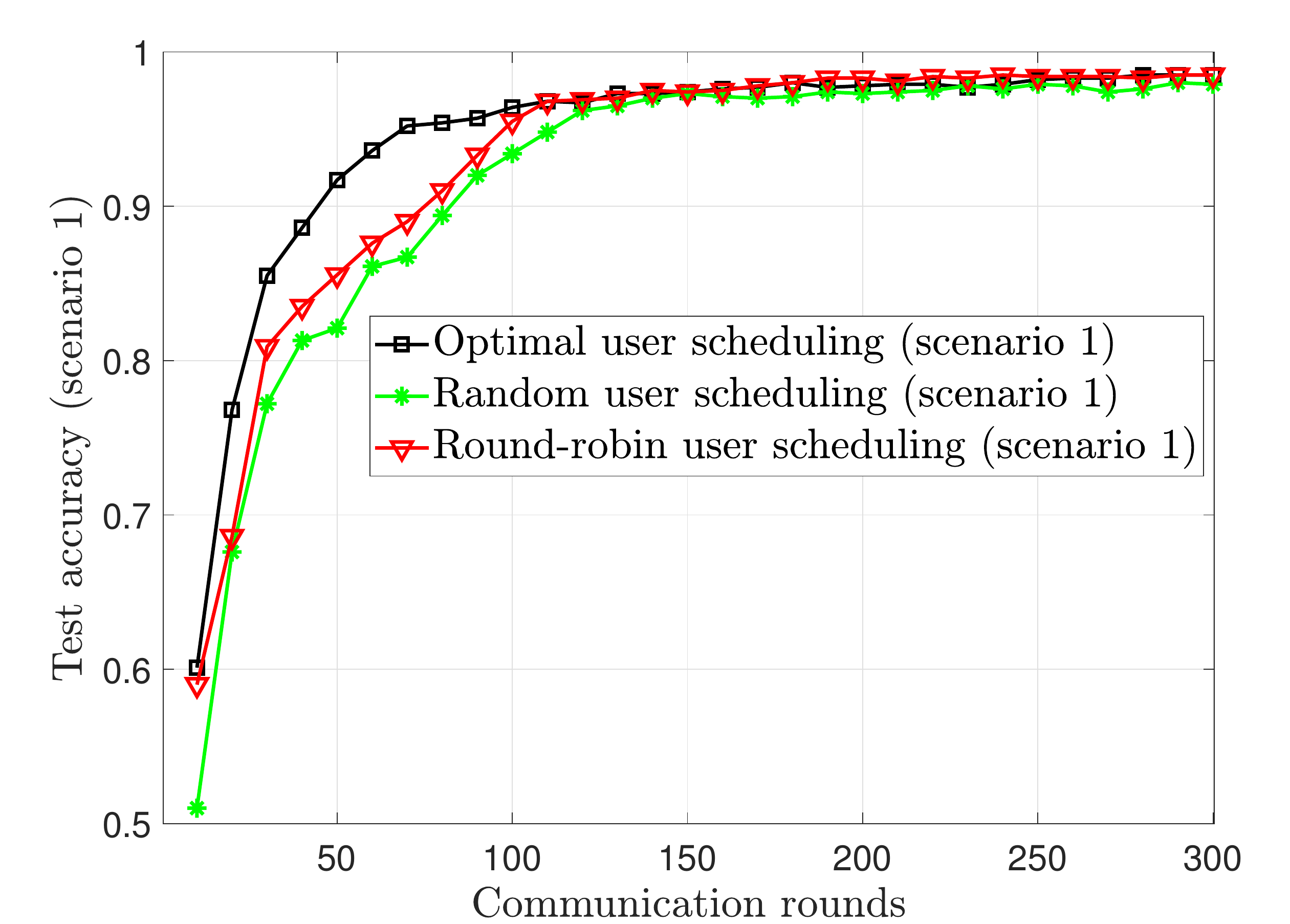}}
	\caption{Test accuracy of the trained CNNs with different scheduling schemes considering a SINR threshold  $\gamma_{th}=25$ dB.}
	\label{fig4}
\end{figure}
Next, the performance of the optimal user scheduling algorithm is evaluated, as shown in Fig.~\ref{fig4}, by considering the ML model known as convolutional neural network (CNN) and an SINR threshold  $\gamma_{th}=25$ dB. We consider CNN with 3 layers and 8 filters each followed by a dense layer of 16 neurons. Similar to the first experiment, a two-class classification task to recognize digits 0 and 8 for each UE with 10 training samples from the MNIST dataset, is considered. This ML model is tested every 10 communications rounds over 100 test samples with a learning rate 0.00006. The optimal scheduling algorithm converges faster than the random and round-robin schemes due to the efficient management of energy and interference. The proposed algorithm reaches a steady state in 100 training steps. However, it requires 150 training steps for the conventional FL systems to reach the steady state. The proposed algorithm increases the number of successful global aggregations. However, the performance gap disappears after a certain number of communications rounds since the difference in terms of successful aggregations is limited when the number of devices is small.
Fig.~\ref{fig4e} shows the impact of the energy arrival rate on FL performance by training CNN. Clearly, the increase of energy arrival rate allows to increase the number of successful global aggregations which highly improve the test accuracy of the trained model.
\begin{figure}[h]
	\centerline{\includegraphics [width=1.05\columnwidth]  {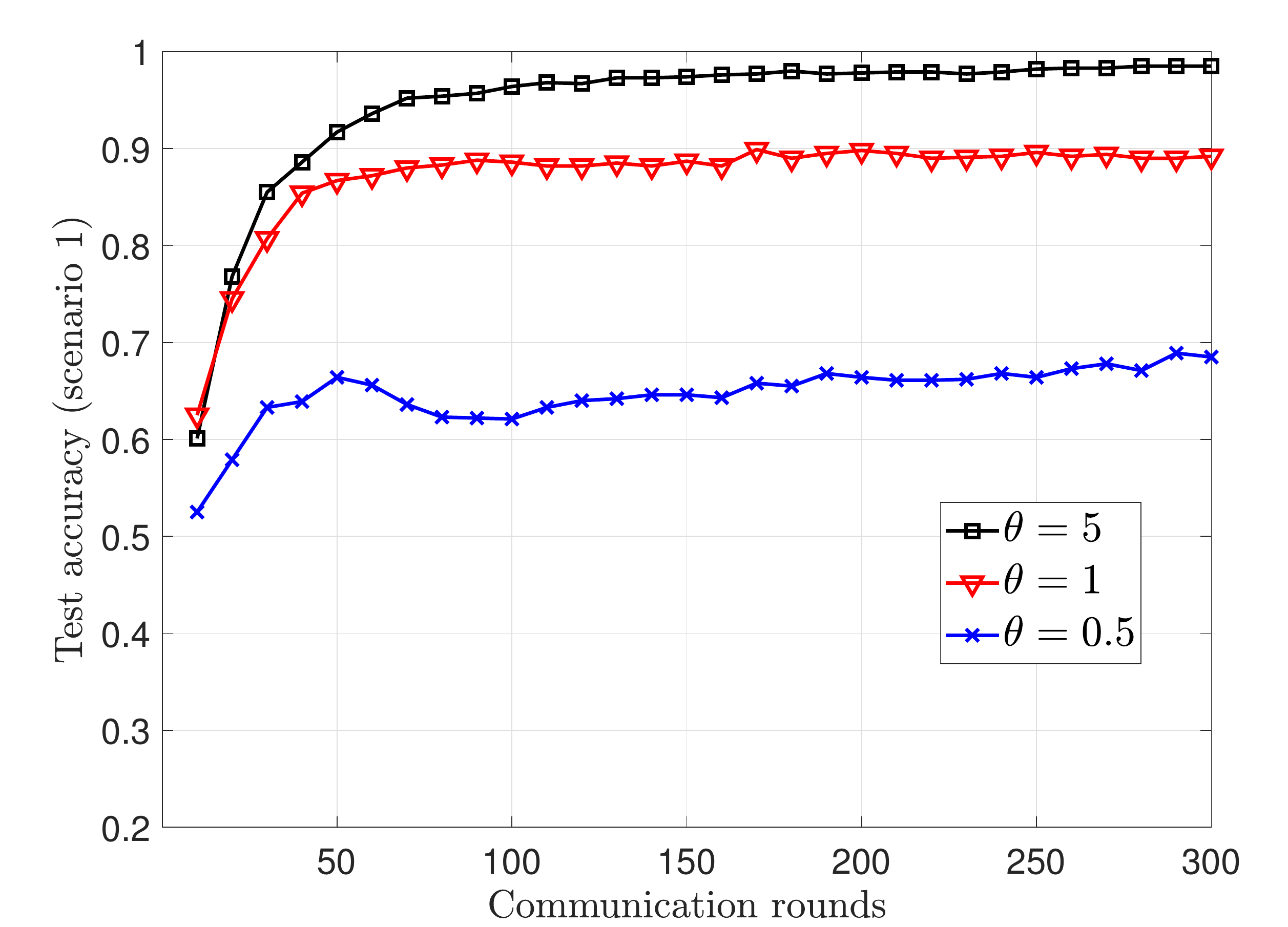}}
	\caption{Impact of the energy arrival rate $\theta$ on test accuracy of the trained CNNs with optimal user scheduling considering a SINR threshold  $\gamma_{th}=25$ dB.}
	\label{fig4e}
\end{figure}
\begin{figure}[h]
	\centerline{\includegraphics [width=1.05\columnwidth]  {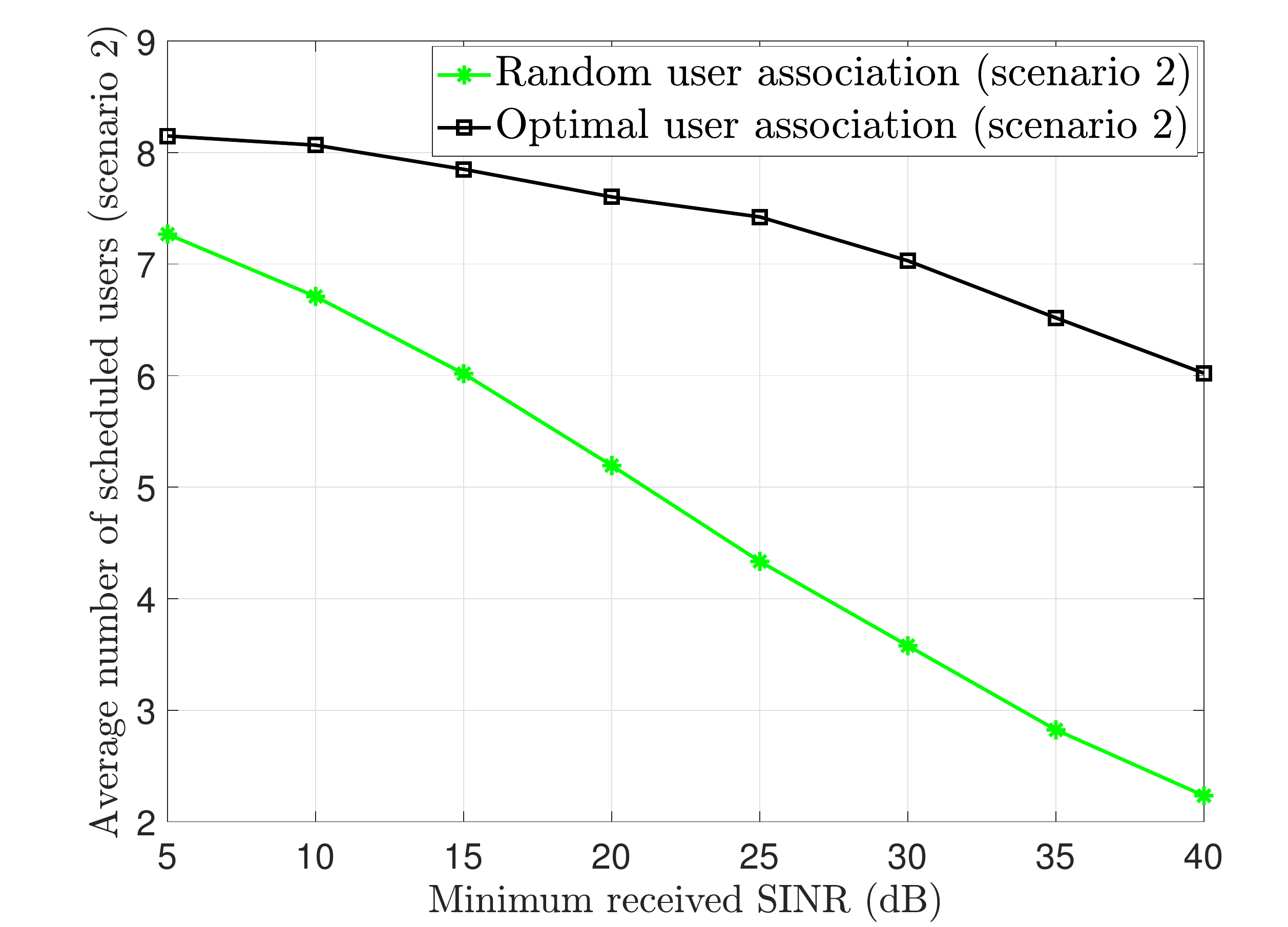}}
	\caption{Average number of scheduled users as a function of the SINR with different user association schemes.}
	\label{fig5}
\end{figure}

\begin{figure}[h]
	\centerline{\includegraphics [width=1.05\columnwidth]  {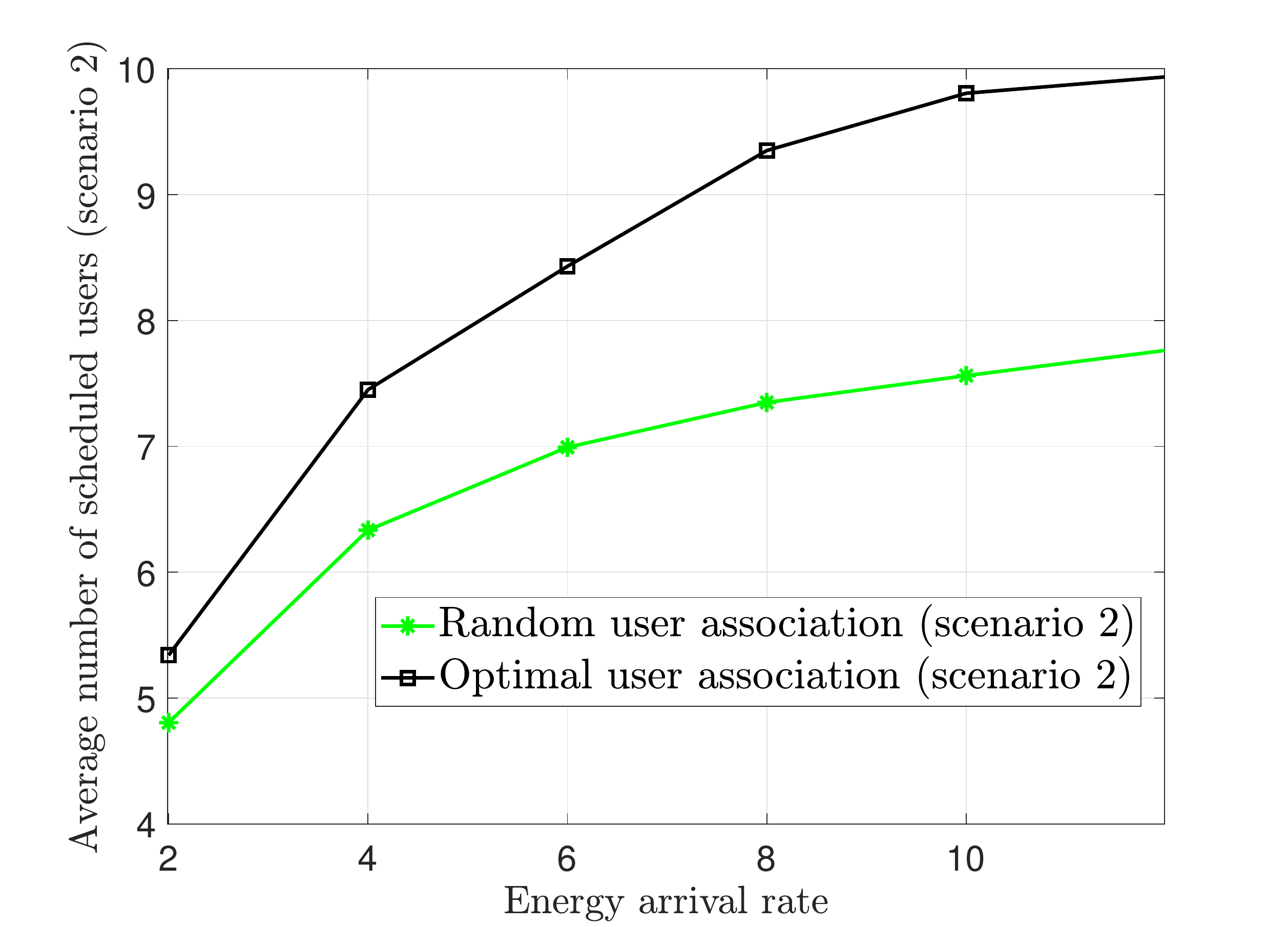}}
	\caption{Average number of scheduled users as a function of the energy arrival rate $\theta$ with different user association schemes considering a SINR threshold  $\gamma_{th}=10$ dB.}
	\label{fig5e}
\end{figure}

In Fig.~\ref{fig5}, a multiple BS system is investigated assuming $S=4$ by showing the performance of the optimal user association in terms of the average number of scheduled users versus the minimum received SINR. It is can be seen that the average number of scheduled users is lower when the QoS constraints are more stringent. Also, the optimal user association allows to admit more links compared to the random user association. The performance gap between the two user association schemes increases for high SINR threshold due to the efficient management of the harvested energy and interference. Fig.~\ref{fig5e} shows the performance of the user association schemes in terms of the average number of scheduled users versus the minimum received SINR. It is clear that the number of scheduled users increases when the energy arrival rate increases. However, the performance tends to saturate for the random association scheme due to interference.\\
\begin{figure}[h]
	\centerline{\includegraphics [width=1.05\columnwidth]  {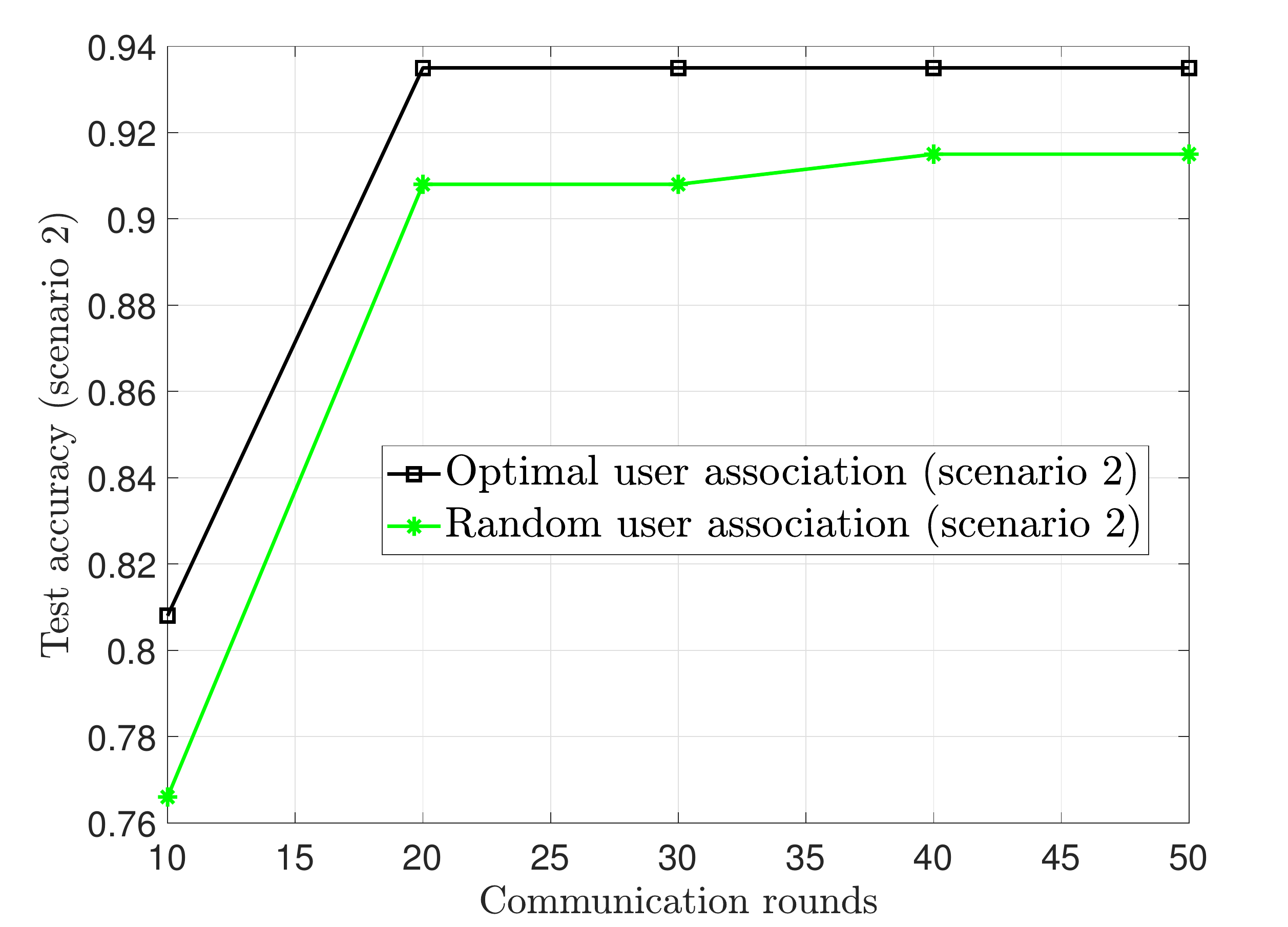}}
	\caption{Test accuracy of the trained SVM with different association schemes considering a SINR threshold  $\gamma_{th}=25$ dB.}
	\label{fig6}
\end{figure}
In Fig.~\ref{fig6}, the effect of user association schemes on FL performance is shown. The performance of the optimal user association is evaluated in FL over wireless multiple BSs system considering SVM and assuming $S=4$ and SINR threshold  $\gamma_{th}=25$ dB. A two-class classification task to recognize digits 0 and 8 for each UE with 2 training samples from the MNIST dataset, is considered. This ML model is tested every 10 communications rounds over 100 test samples with a learning rate 0.00001. It is clear that the optimal user association enhances the FL performance in terms of accuracy compared to random association, since it increases the successful global aggregations. This mainly is due to fact that optimal user association scheme can tackle more efficiently the interference and energy constraints.
\begin{figure}[h]
	\centerline{\includegraphics [width=1.05\columnwidth]  {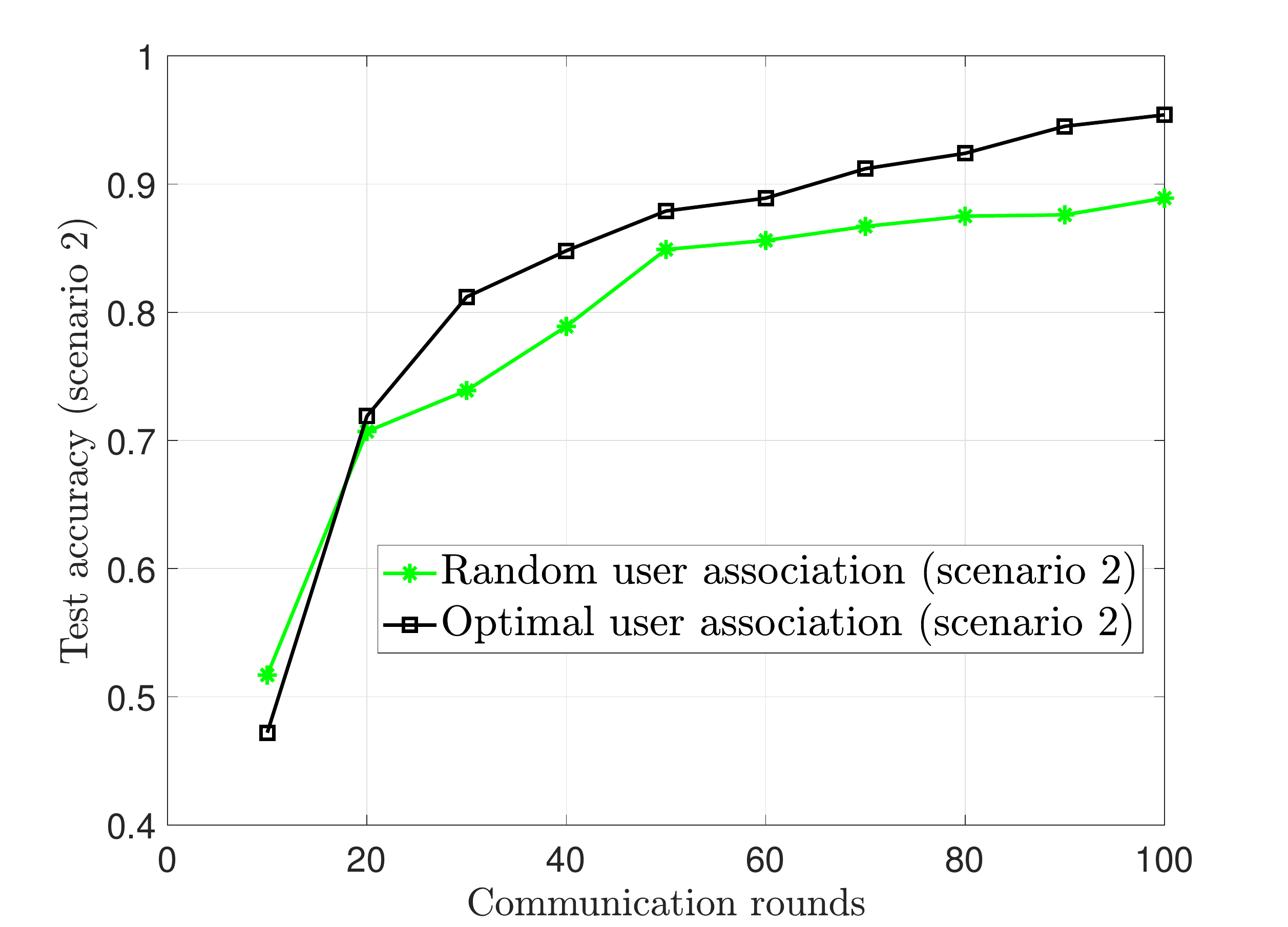}}
	\caption{Test accuracy of the trained CNN with different association schemes considering a SINR threshold  $\gamma_{th}=30$ dB.}
	\label{fig7}
\end{figure}
Also, the multiple BS system is investigated in Fig.~\ref{fig7} and the effect of user association schemes on FL performance is shown assuming CNN. A two-class classification task to recognize digits 0 and 8 for each UE with 10 training samples from the MNIST dataset, is considered. This ML model is tested every 10 communications rounds over 100 test samples with a learning rate 0.00006. It is clear that the proposed user association scheme in FL over wireless systems allows to increase the total number of aggregations which allows to enhance the FL performance in terms of accuracy and time convergence. The optimal user association algorithm converges faster than the random scheme due to the efficient management of energy and interference.

\begin{figure*}[t]	
	\begin{equation}
	\label{eq:14}
	\begin{aligned}
	\mathbb{E}\left(\|\boldsymbol{o}\|^{2}\right) &=\mathbb{E}\left(\left\|\nabla F\left(\boldsymbol{q}_{i}\right)-\frac{\sum\limits_{k=1}^{K} \sum\limits_{m=1}^{M_{k}} \chi_k(i) \nabla f\left(\boldsymbol{q}, \boldsymbol{x}_{k,m}, y_{k,m}\right) }{\sum\limits_{k=1}^{K} M_{k} \chi_k(i) }\right\|^{2}\right) \\
	&=\mathbb{E}\left(\left\|-\frac{(K-\sum\limits_{k=1}^{K} M_{k} \chi_k(i)) \sum\limits_{k \in \mathcal{N}_{2}(i)} \sum\limits_{m=1}^{M_{k}} \nabla f\left(\boldsymbol{q}, \boldsymbol{x}_{k,m}, y_{k,m}\right)}{K \sum\limits_{k=1}^{K} M_{k} \chi_k(i)} \right. \right.
	\left.\left.+\frac{\sum\limits_{k \in \mathcal{N}_{1}(i)} \sum\limits_{m=1}^{M_{k}} \nabla f\left(\boldsymbol{q}, \boldsymbol{x}_{k,m}, y_{k,m}\right) }{K}\right\|^{2}\right) \\
	& \leq \mathbb{E}\left(-\frac{(K-\sum\limits_{k=1}^{K} M_{k} \chi_k(i)) \sum\limits_{k \in \mathcal{N}_{2}(i)} \sum\limits_{m=1}^{M_{k}} \left\| \nabla f\left(\boldsymbol{q}, \boldsymbol{x}_{k,m}, y_{k,m}\right)\right\|}{K \sum\limits_{k=1}^{K} M_{k} \chi_k(i)} \right.
	\left.+\frac{\sum\limits_{k \in \mathcal{N}_{1}(i)} \sum\limits_{m=1}^{M_{k}} \left\| \nabla f\left(\boldsymbol{q}, \boldsymbol{x}_{k,m}, y_{k,m}\right)\right\| }{K}\right)^{2}.
	\end{aligned}
	\end{equation}
	\hrule
\end{figure*}

\section{conclusion}
This paper has investigated an energy-efficient FL over wireless system by equipping the BS with massive MIMO and powering the users by harvested energy sources. A minimization problem involving a global loss function subject to a quality of service constraint per user and energy availability constraints, has been formulated. We have investigated the impact of wireless parameters on Fl performance and derived the relationship between the FL convergence rate and user scheduling. The optimal user scheduling that minimizes the FL convergence rate is derived. Moreover, the multiple BSs case has been investigated by optimizing the optimal user association in FL over wireless systems using branch and bound.

\section*{Appendix A \\ Proof of Theorem 1}
The following inequality shows the relationship between the convergence rate of FL model and the wireless parameters in~\cite{us9}. It is adapted to our system model as (\ref{eq:14}), where $\mathcal{N}_{1}(i)=\{k \in \{1,\ldots,K\}  \mid \chi_k(i)=1\}$ is the set of scheduled users at frame $i$ and $\mathcal{N}_{2}(i)=\{k \in \{1,\ldots,K\}  \mid \chi_k(i)=0\}$ is the set of unscheduled users at frame $i$. Hence, the following inequality holds:
\begin{equation}
	\label{eq:15}
	\begin{aligned}
	&\mathbb{E}\left[F\left(\boldsymbol{q}_{i+1}\right)-F\left(\boldsymbol{q}^{*}\right)\right]  \leq \frac{2 \zeta_{1}}{V M} \sum_{k=1}^{K} M_{k}\left(1-\chi_k(i)\right) 
	+(1-  \\
	& \frac{\mu}{V}+\frac{4 \mu \zeta_{2}}{V M} \sum_{k=1}^{K} M_{k}\left(1-\chi_k(i)\right)) \mathbb{E}\left(F\left(\boldsymbol{q}_{i}\right)-F\left(\boldsymbol{q}^{*}\right)\right).
	\end{aligned}
\end{equation}
Let~$u_i=\mathbb{E}[F(\boldsymbol{q}_{i})-F(\boldsymbol{q}^{*})],a_i=\frac{2 \zeta_{1}}{V M} \sum_{k=1}^{K} M_{k}(1-\chi_k(i))$, and
 $b_i=1-\frac{\mu}{V}+\frac{4 \mu \zeta_{2}}{V M} \sum_{k=1}^{K} M_{k}\left(1-\chi_k(i)\right)$. Hence, the following inequality holds:
\begin{equation}
	\label{eq:16}
	u_{i+1} \leq a_i +b_i u_i.
\end{equation}
Now, the following inequality is recursively proven:
\begin{equation}
	\label{eq:17}
	u_{i+1}  \leq \sum_{t=0}^{i} \left(\prod_{j=t+1}^{i} b_j\right) a_t   + u_0 \prod_{t=0}^{i} b_t.
\end{equation}
For $i=0$, the previous inequality is true as we have
\begin{equation}
	\label{eq:18}
	u_{1} \leq a_0 +b_0 u_0.
\end{equation}
Let assume that the inequality is true for $i+1$ and prove it for $i+2$ as

\begin{equation}
	\label{eq:19}
	\begin{aligned}
	u_{i+2} & \leq a_{i+1} +b_{i+1} u_{i+1}\\
	&  \leq a_{i+1} + b_{i+1} \left( \sum_{t=0}^{i} \left(\prod_{j=t+1}^{i} b_j\right) a_t   + u_0 \prod_{t=0}^{i} b_t \right) \\
	& \leq  a_{i+1} +  \sum_{t=0}^{i} \left(\prod_{j=t+1}^{i+1} b_j\right) a_t    + u_0 \prod_{t=0}^{i+1} b_t \\
	& \leq \sum_{t=0}^{i+1} \left(\prod_{j=t+1}^{i+1} b_j\right) a_t    + u_0. \prod_{t=0}^{i+1} b_t.
	\end{aligned}
\end{equation}
This completes the proof.

\section*{Acknowledgment}
This research was sponsored in part by the TÜB˙ITAK—QNRF Joint Funding Program under Grant AICC03-0324-200005 from the Scientific and Technological Research Council of Turkey and Qatar National Research Fund (QNRF) and in part by the U.S. National Science Foundation under Grant CCF-1908308.

\balance

\end{document}